\documentclass[letterpaper, 10 pt, conference]{ieeeconf}  

\DeclareUnicodeCharacter{2009}{\,}

\IEEEoverridecommandlockouts                              
\usepackage{hyperref}
\usepackage{booktabs}
\usepackage{graphicx}

\title{\LARGE \bf
Reimagining Assistive Walkers: An Exploration of Challenges and Preferences in Older Adults
}

\author{Victory A. Aruona$^{1}$, Sergio D. Sierra M.$^{1}$, Nigel Harris$^{1}$, Marcela Munera$^{1}$, Carlos A. Cifuentes$^{1}$
\thanks{*This work was supported by the UKRI Medical Research Council [grant number MR/Y010620/1], EPSRC, and FARSCOPE CDT. }
\thanks{$^{1}$All authors are with the Bristol Robotics Laboratory ,
        University of the West of England, Bristol, UK. victory2.aruona@live.uwe.ac.uk,
        \{sergio.sierramarin, nigel2.harris, marcela.munera, carlos.cifuentes\}\@uwe.ac.uk}%
}

\begin{document}

\maketitle
\thispagestyle{empty}
\pagestyle{empty}

\begin{abstract}

The well-being of older adults relies significantly on maintaining balance and mobility. As physical ability declines, older adults often accept the need for assistive devices. However, existing walkers frequently fail to consider user preferences, leading to perceptions of imposition and reduced acceptance. This research explores the challenges faced by older adults, caregivers, and healthcare professionals when using walkers, assesses their perceptions, and identifies their needs and preferences. A holistic approach was employed, using tailored perception questionnaires for older adults (24 participants), caregivers (30 participants), and healthcare professionals (27 participants), all of whom completed the survey. Over 50\% of caregivers and healthcare professionals displayed good knowledge, positive attitudes, and effective practices regarding walkers. However, over 30\% of participants perceived current designs as fall risks, citing the need for significant upper body strength, potentially affecting safety and movement. More than 50\% highlighted the importance of incorporating fall detection, ergonomic designs, noise reduction, and walker ramps to better meet user needs and preferences.

\end{abstract}

\section{Introduction}

Mobility is fundamental for performing essential tasks and actively participating in societal life, allowing older adults to live more independently \cite{NationalInstitute2020}. In 2017, the global population over 60 years old was 962 million, and it is forecast to double again by 2050, reaching 2.1 billion \cite{UnitedNations2017}. With the rising life expectancy among older adults, more of them are facing mobility issues that may require assistance with daily tasks \cite{Jaul2017, Abdi2020}. 

Loss of physical mobility makes full participation in desired activities more complicated and could prevent social involvement \cite{Cowan2012}. The impact of limited mobility on mental health, the risk of falls, disability, hospitalization, and mortality rates cannot be understated, as indicated by studies \cite{Lee2023, Davis2015, Musich2018}. Therefore, mobility plays a pivotal role in fostering independence, social engagement, and overall well-being as individuals age \cite{Pantelaki2021, Rosso2013}. 

Assistive technologies can enhance autonomy, safety, and quality of life of the ageing population \cite{Baccman2019}. Among these technologies, assistive walkers stand out for their potential to improve ambulation and promote physical activity \cite{Fomiatti2013}. For instance, walkers are used indoors by 22\% and outdoors by 44\% of older adults in the UK \cite{Lofqvist2007}. This category of devices includes canes, standard walkers, and wheeled walkers. 

While mobility devices have been shown to boost independence, foster social participation, and improve the quality of life in older adults \cite{Boerema2017}, the use of assistive walkers presents its own set of challenges. Despite some studies suggesting that walking aids may prevent falls in older individuals \cite{Graafmans2003}, the association remains poorly understood \cite{Stevens2009}. This is because walking aids can alter gait patterns by preventing normal arm swing, changing posture, decreasing gait speed, and increasing standing time \cite{Liu2009}. Difficulty in navigating stairs, destabilizing biomechanical effects, upper limb pain, and an increased risk of falling are among the potential drawbacks \cite{Bateni2005, Bradley2011}. 

Moreover, assistive walkers frequently fail to satisfy the needs of older adults, and this misalignment between their needs and the adoption of these technologies stymies their capacity to achieve independence \cite{Soar2020}. For instance, canes and wheeled walkers may fall short in providing adequate support during essential sit-to-stand transfers that are crucial for daily activities \cite{Schmidt2017}. Several studies have examined older adults' experiences and satisfaction with mobility aids. The study in \cite{Samuelsson2008} found high user satisfaction with rollators and manual wheelchairs. However, differences were noted in how users assessed the usefulness and characteristics of each device. The systematic review in \cite{Schmucker2025} analysed 45 studies to assess the usability of rollators, focusing on aspects such as ease of use, comfort, and safety. Using a deductive approach, they found that while these factors were frequently discussed, areas like repairs, servicing, and durability received less attention. The study highlights the lack of dedicated research on rollators' usability and calls for future user-centred, participatory studies to improve satisfaction and evaluation in assistive mobility device design.

Thus, understanding the challenges faced by older adults using assistive walkers, analysing their perspectives, and examining the insights of healthcare professionals and caregivers are essential steps towards the development of more effective, user-centric solutions. This study explores the challenges encountered by older adults using assistive walkers to enhance gait activities. We aim to analyse older adults' perspectives on the features they desire in an ideal assistive walker, aligning future advancements with their unique requirements and preferences. Additionally, we explore the perspectives of healthcare professionals and caregivers, who possess valuable first-hand experience with the mobility difficulties faced by older adults. This approach aims to bridge the gap between existing assistive walker technologies and the evolving needs of the ageing population.

\section{Materials and Methods}
This section outlines the perception questionnaires for older adults, healthcare professionals, and caregivers. 

\subsection{Demographics}
Information about the older adults' demographics was recorded (See Table \ref{tab:demographics}). The information from healthcare professionals and caregivers, including job titles and experience, was included in the questionnaire.

\begin{table}[t]
\centering
\caption{Characteristics of the older adults}
\label{tab:demographics}
\resizebox{\columnwidth}{!}{%
\begin{tabular}{@{}ll@{}}
\toprule
\textbf{Participants}                         & n = 24, 6 Male, 18 Female \\ \midrule
Age Mean (SD)                                 & 88.67 (6.64) years \\
Weight Mean (SD)                              & 75.98 (6.90) kg    \\ \midrule
\textbf{Health conditions}                    & \textbf{}          \\ \midrule
Vision Problems                               & 79.0 \%            \\
Hearing problems                              & 58.0 \%            \\
Mobility difficulty                           & 92.0 \%            \\ \midrule
\textbf{Health conditions affecting mobility} &                    \\ \midrule
Arthritis                                     & 42.0 \%            \\
Osteoporosis                                  & 5.0 \%             \\
Parkinson disease                             & 8.0 \%             \\
Paralysis                                     & 8.0 \%             \\
Knee and hip condition                        & 42.0 \%            \\
Stroke                                        & 8.0 \%             \\
\bottomrule
\end{tabular}%
}
\end{table}

\subsection{Knowledge, Attitude, and Practice (KAP)}
According to \cite{Andrade2020}, the KAP standardized questionnaire measures a population's knowledge, attitudes, and practices, identifying the initial misconceptions, beliefs, and behaviours in relation to a specific health-related topic. This questionnaire also provides information on needs, issues, and barriers related to technology. The questionnaire aimed to gather information from health professionals and caregivers regarding their knowledge base, overall perception, and everyday activities using assistive walkers. Questions were phrased as statements and graded on a 5-point Likert scale ranging from -2 (strongly disagree) to 2 (strongly agree). 

\subsection{Closed-Ended Questions}
The functionality and limitations of conventional assistive walkers (CAWs) were assessed using a series of statements that were developed following a literature review. A 5-point Likert scale was used. Questions aimed at assessing the walker's effectiveness, safety, comfort, stability, and support for daily activities, mobility, and transitions both indoors and outdoors. Questions also included perceptions of fall risk, navigation support, user-friendliness, complexity, movement limitations, and physical effort required.

\subsection{Open-Ended Questions}
The open-ended questions evaluated the features of CAWs that older adults consider necessary. These questions also collected suggestions, perceived drawbacks and benefits, ideas for improvement, and overall perceptions from caregivers and healthcare professionals. Questions assessed preferences, concerns, and ideal walker features. Moreover, questions inquired about perceptions of familiar walkers, emphasizing limitations faced and proposed enhancements. 

\subsection{Research setting and protocol}
The study was conducted in two private residential care homes in south-west England, selected for their high number of older adults using assistive walkers and participants’ cognitive abilities. Caregivers and physiotherapists with relevant experience were recruited from these homes, while additional caregivers and healthcare professionals were recruited from other facilities and institutions. All participants provided informed consent after reviewing the consent statement. They then submitted demographic information and completed the survey. Ethical approval was granted by the University of West of England's Ethics Committee.

\subsection{Participant Recruitment}
The study included older adults selected using non-random convenience sampling. Medical information regarding memory function was reviewed with the physiotherapist prior to interviews. Inclusion criteria were: (1) aged 75 or older, (2) living in residential or home care settings, (3) requiring mobility assistance and having visited a hospital for related issues, (4) relying on a walker for daily transportation, or (5) prescribed a walker. Exclusion criteria included: (1) inability to walk and (2) cognitive impairment.

The study also included physiotherapists, occupational therapists, and general practitioners with experience working with older adults, as well as caregivers providing support to individuals using assistive walkers. Inclusion criteria were: (1) providing care to older adults, (2) working in care homes, and (3) at least six months of experience in elderly care. Exclusion criteria included: (1) no experience attending to older adults.

\subsection{Data Analysis}
The study employed a mixed methods approach to explore participants’ perceptions towards the use of CAWs among older adults. The primary focus was qualitative, aiming to capture the depth and richness of participants’ experiences and preferences through open-ended questions. To complement the qualitative insights, a Likert scale was utilized to quantify the levels of agreement with specific statements related to assistive walkers. Responses were summarized by examining the percentage of agreement, neutrality, and disagreement, providing a comprehensive view of participants attitude and preference. 

\section{Results}
Data were collected from a total of 81 participants (24 older adults, 30 caregivers and 27 clinicians). The outcomes from the KAP survey, the closed-ended and the open-ended questions, are discussed as follows. 

\subsection{Older Adults}
\begin{figure}[t]
     \centering
     \includegraphics[scale=0.6]{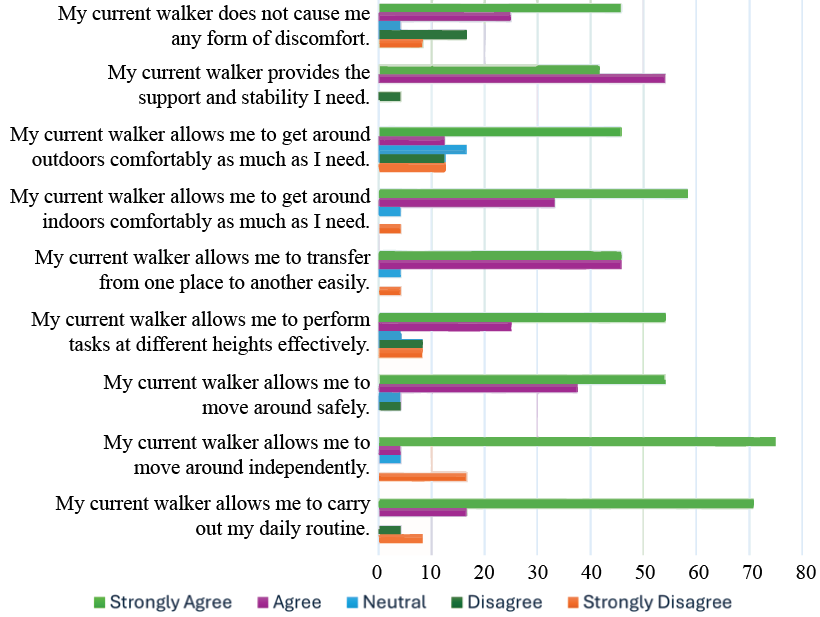}
     \caption{Older adult's opinions on the functionalities of the assistive walkers}
     \label{fig:functionalities}
 \end{figure}

A total of 24 older adults took part in the study. Table \ref{tab:demographics} details their characteristics; most participants had a visual or hearing impairment, with arthritis and/or hip or knee problems affecting their mobility. Figure \ref{fig:functionalities} presents older adults' opinions on the assistive walker's functionalities. Most older adults perceive their current walkers as effective tools for maintaining daily routines and ensuring independent mobility, with 88\% and 96\% respectively agreeing with these statements. However, opinions on specific aspects of walker use are more varied. While 79\% feel safe using their walkers, concerns persist regarding stability and comfort, with 21\% questioning safety and 25\% experiencing discomfort. Walkers are generally considered effective for indoor use, as 91\% of users report comfort within their homes. Outdoor usability is less favourable, with only 59\% finding their walkers suitable for external environments. Additionally, while 79\% agree that their walkers help in performing tasks at different heights, and 92\% find them useful for easy transfers, there are still notable percentages of users who disagree. 

\subsection{Healthcare professionals and Caregivers}
A total of 30 caregivers and 27 healthcare professionals took part in the study. Regarding work experience in elderly care, approximately 60\% of caregivers had more than 2 years' experience and 78\% of healthcare professionals had more than 2 years' experience. Regarding the number of older adults being cared for, 64\% of caregivers cared for more than 10 older adults while 59\% of clinicians attended to more than 10 older adults.

\begin{figure}[t]
    \centering
    \includegraphics[scale=0.6]{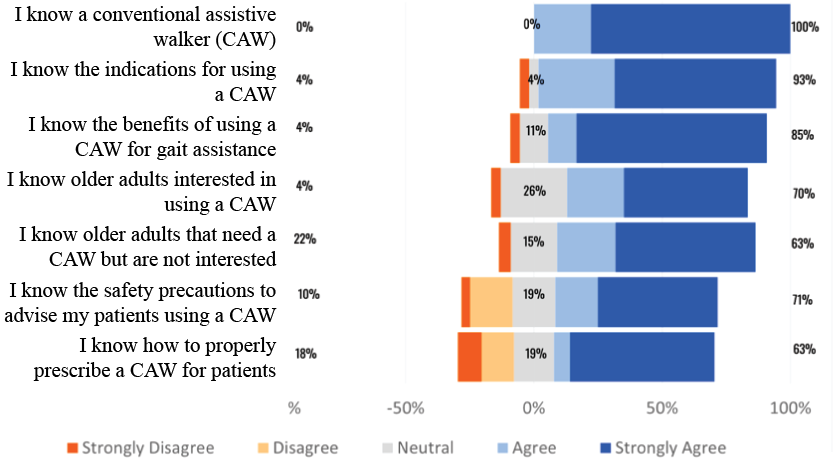}
    \caption{Knowledge of healthcare professionals regarding the use of an assistive walker. Percentages show total disagreement, neutral and agreement.}
    \label{fig:know}
\end{figure}

\begin{figure}[t]
    \centering
    \includegraphics[scale=0.53]{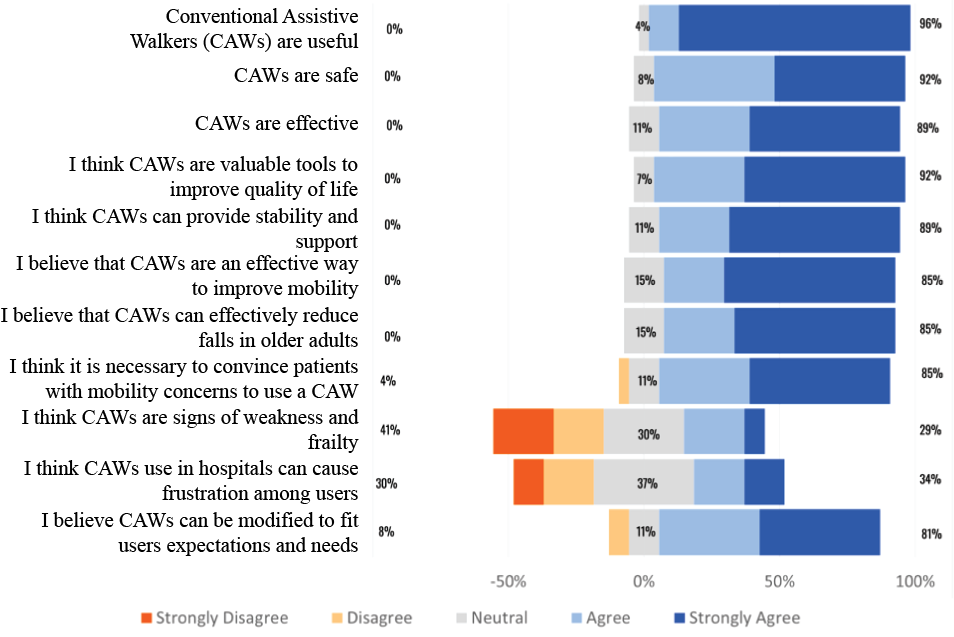}
    \caption{Attitude of healthcare professionals regarding the use of an assistive walker. Percentages show total disagreement, neutral and agreement.}
    \label{fig:attitudes}
\end{figure}

\begin{figure}
    \centering
    \includegraphics[scale=0.5]{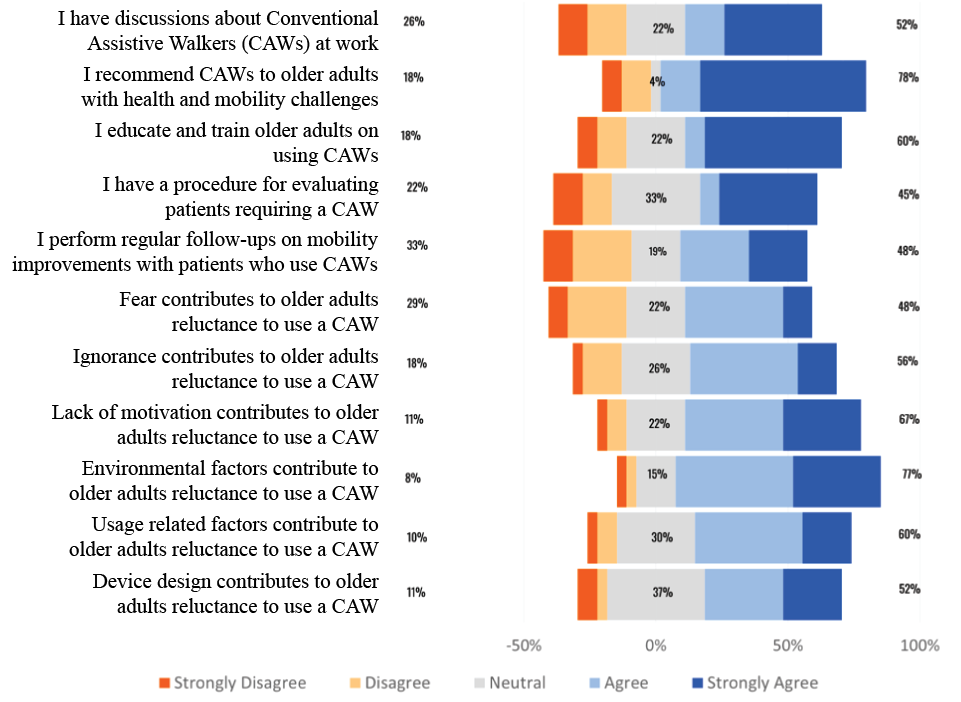}
    \caption{Practice of healthcare professionals regarding the use of an assistive walker. Percentages show total disagreement, neutral and agreement.}
    \label{fig:practice}
\end{figure}

Figure \ref{fig:know}, Figure \ref{fig:attitudes} and Figure \ref{fig:practice} present the level of agreement of healthcare professionals to the KAP questions. As expected, Figure \ref{fig:know} shows that health professionals have a high level of knowledge, with 50\% and above agreement with the statements in this category, but with some uncertainty around safety precautions and prescribing. Figure \ref{fig:attitudes} shows health professionals’ attitudes, which are overwhelmingly positive, with 75\% and above agreement with most statements in this category, reflecting a positive attitude to the use of assistive walkers. However, there is 50\% neutral agreement with statements “I think assistive walkers serve as a sign of weakness and frailty" and “I think CAWs used in hospital settings can cause frustration among users”. Figure \ref{fig:practice} explores practice, there is 50\% and above agreement with statements in this category showing good practice by these respondents on the usage of assistive walkers. However, there is 50\% neutral agreement with statements “I have a procedure for evaluating patients requiring a CAW”, “I perform regular follow up on mobility improvements with patients who use a CAW”, and “I have observed older adults are reluctant to use a CAW due to fear”. 

\begin{figure}[t]
    \centering
    \includegraphics[scale=0.6]{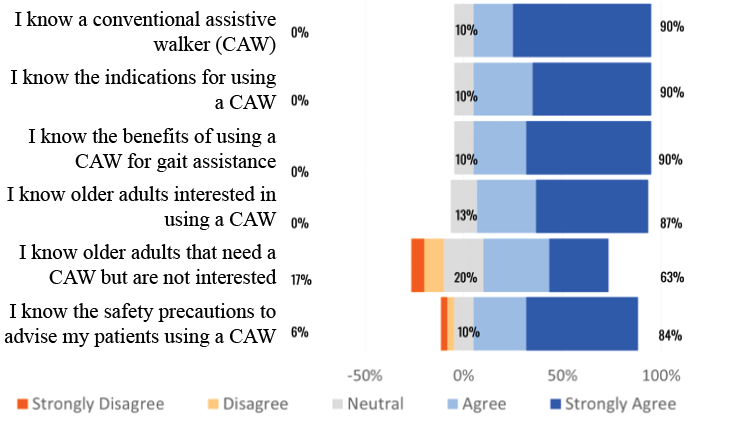}
    \caption{Knowledge of caregivers regarding the use of an assistive walker. Percentages show total disagreement, neutral and agreement.}
    \label{fig:know_care}
\end{figure}

\begin{figure}[t]
    \centering
    \includegraphics[scale=0.5]{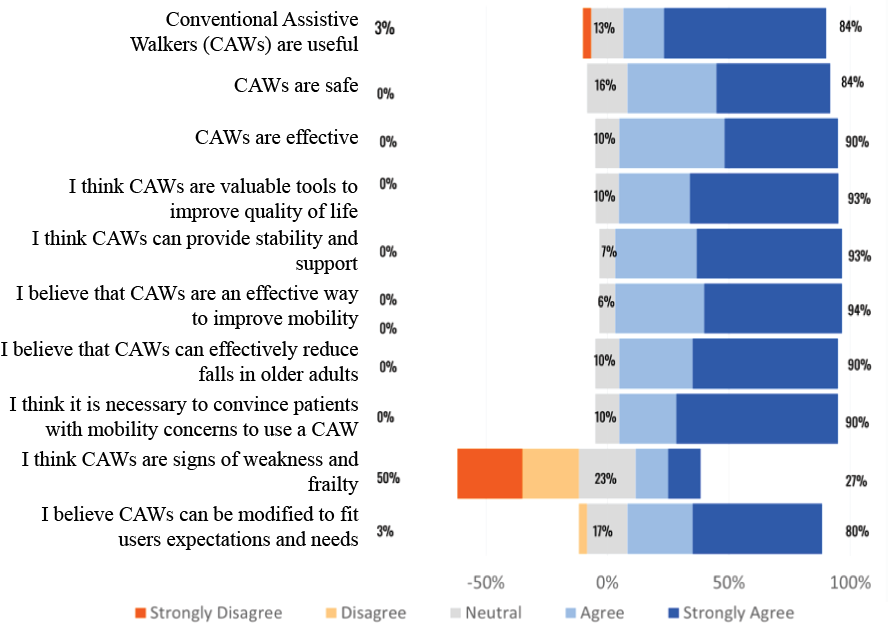}
    \caption{Attitude of caregivers regarding the use of an assistive walker. Percentages show total disagreement, neutral and agreement.}
    \label{fig:attitude_care}
\end{figure}

\begin{figure}
    \centering
    \includegraphics[scale=0.5]{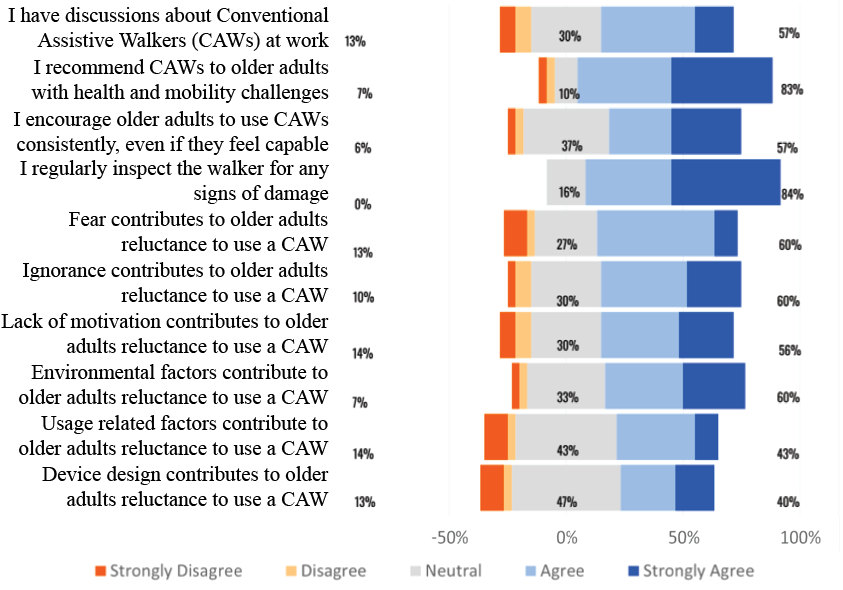}
    \caption{Practice of caregivers regarding the use of an assistive walker. Percentages show total disagreement, neutral and agreement.}
    \label{fig:practice_care}
\end{figure}

Figure \ref{fig:know_care} shows there is 50\% and above agreement to the statements in knowledge, highlighting a strong foundational knowledge of caregivers regarding using an assistive walker. Figure \ref{fig:attitude_care} shows there is 100\% agreement with most statements in attitude, reflecting a positive attitude to the use of assistive walkers. However, there was a 25\% mixed perception with the statement “I think assistive walkers serve as a sign of weakness and frailty". Figure \ref{fig:practice_care} shows 75\% agreement with statements in this practice showing good practice by these respondents on the usage of assistive walkers. However, 50\% of respondents were neutral on whether older adults avoid CAWs due to usage factors or design.

\subsection{Limitations of assistive walkers}
The analysis of perceptions revealed diverse concerns and satisfaction levels. While 55\% of older adults feel safe using walkers, 37\% of both older adults and caregivers, along with 33\% of healthcare professionals, perceive walkers as fall risks due to stability issues. A major gap identified is the lack of navigation systems, highlighted by 96\% of older adults, suggesting a need for technological enhancements. Usability is generally rated positively, with 80\% of older adults, 60\% of caregivers, and 59\% of healthcare professionals finding walkers easy to use, though some complexity remains for 21\% of older adults. Movement restrictions are less of a concern for older adults, but some caregivers and healthcare professionals disagree, indicating design adaptability needs. The physical effort required to use walkers is noted by 42\% of older adults and 40\% of caregivers, pointing to the need for ergonomic improvements to enhance safety and ease of use.

\subsection{Perception of Older adults, Caregivers and Healthcare professionals towards the use of an assistive walker}
The result shows consistent preference of assistive walkers across all groups. Common challenges included fear of falling, difficulty navigating tight spaces, and usage challenges relating to size, noise, and brakes. To mitigate this, more than 50\% of each group recommended the need for innovative features such as navigation systems, ergonomic designs, and threshold ramps. Above 50\% of caregivers and healthcare professionals also noted the need for additional safety and convenience features including alarm buttons, user-friendly brakes, and fall detection systems. Additionally, over 50\% of healthcare professionals advised the inclusion of noise reduction features and enhanced device designs\footnote{See complementary material with perception answers: \href{https://figshare.com/articles/dataset/Data_for_Reimagining_Assistive_Walkers_An_Exploration_of_Challenges_and_Preferences_in_Older_Adults/27985727?file=51040010}{here}}.

\section{Discussion}
Most caregivers and healthcare professionals demonstrated a strong understanding, positive outlook, and proficient usage of these walkers, emphasizing their significance for older adults. The respondent groups expressed concerns about the frequent falling incidents related to the walker usage, with older adults sharing their experiences of falling while transferring and navigating with the walker. Older adults and caregivers blamed walker deficiencies, such as wheel issues and poor brakes. However, healthcare professionals attributed falls to poor balance and cognitive impairments, this was particularly problematic in cognitively impaired patients, as also stated in previous studies by \cite{ONeill2010, Nilsson2011}. This implies there would be benefit in incorporating cognitive assistance features into these walkers to facilitate smoother transfers.

Healthcare professionals reported walkers to be heavy, requiring so much strength and effort, further exacerbating the issues and increasing the risk of falling which supports a study by \cite{Sakano2023}. Both older adults and caregivers reported that the walkers were not enough to support their weight, causing discomfort, pain, and psychological distress. This statement is consistent with \cite{Mali2023} indicating that individuals with a much higher weight find it difficult to move around comfortably. In addition, caregivers reported that the noise from the walkers, particularly from the wheels and brakes, could cause discomfort, stress, and agitation, leading to reluctance to use the walkers.

All respondents' groups expressed concerns about walkers' limited manoeuvrability in certain situations, like stairs, narrow doorways, and thresholds. The walker's size, including width, length, and height, significantly affects its ability to move effectively in confined spaces \cite{Thies2020}. Older adults reported getting stuck while taking turns or negotiating tight corners, while caregivers confirmed experiencing this with the older adults they care for. This situation can be particularly challenging in places like bathrooms or crowded areas, where precise navigation is crucial for the user's safety and convenience \cite{Nickerson2024, Sloot2022}.

The results from this research have outlined additional features:

\subsection{Smart Design}
Integrating an alarm function allows users to access emergency assistance quickly, especially during outdoor falls where pendant alarms may fail. While studies \cite{Messaoudi2022, Cullen2022} highlight the importance of such features, their feasibility depends on factors such as cost, power consumption, and user familiarity with GPS-enabled devices. Simplified interfaces and low-maintenance designs could enhance adoption among older adults.

Assisting users in navigating unfamiliar environments, detecting obstacles, and guiding posture, this feature could improve walker interaction. However, implementation must balance functionality with ease of use, ensuring it does not increase device complexity or require high technological literacy.

Measuring grip force could support fall risk assessments and intervention. Older adults showed strong interest in this feature (85\% approval). However, maintenance and battery requirements must be considered to ensure long-term usability. Studies \cite{Krafft2022} indicate stability benefits, but further evaluation is needed to confirm cost-effectiveness.

\subsection{Ergonomic Design}
An ergonomically designed walker should accommodate various user heights and body types, promoting optimal posture while reducing physical effort. Features like adjustable size, wheels, and brakes \cite{Wang2013} could enhance stability on uneven terrain. However, adjustable mechanisms must be intuitive to prevent usability barriers.

\subsection{Mechanical Design}
Integrating noise reduction features, such as noise-absorbing materials in wheels and brakes, could dampen vibrations and improve user experience. Additionally, an attachable threshold ramp could aid smoother transitions over door thresholds \cite{Thies2023}. While these features enhance accessibility, factors such as added weight, durability, and maintenance requirements must be considered to avoid potential trade-offs that could deter usage.

\subsection{Limitations}


Older adults and some caregivers preferred in-person surveys, while clinicians and other caregivers preferred virtual formats. This approach, while ensuring broader participation, is not believed to have biased the findings. Older adults were recruited exclusively from two private care homes, while caregivers and healthcare professionals came from these homes and other institutions. Variations in knowledge, attitudes, and practices (KAP) may exist; however, they likely reflect normal professional differences. The use of non-random convenience sampling, based on availability and willingness to participate, and the exclusive focus on care home residents may have introduced bias. Future studies should aim for a broader sample, including older adults living independently and participants from more diverse cultural and socioeconomic backgrounds to enhance generalizability.

The study's applicability is also constrained by its sample size of 81 participants, though this remains one of the largest studies of its kind. Additionally, the questionnaire did not explicitly differentiate between various types of assistive devices (e.g., canes, standard walkers, wheeled walkers), which may have led to overly broad conclusions. Future research should refine the survey design to distinguish between device types and provide more targeted recommendations. Furthermore, our survey design did not incorporate reverse semantic questions, which are often used to assess response reliability. While we mitigated this by structuring the survey clearly and ensuring logical consistency in responses, we acknowledge that incorporating additional validation techniques could strengthen data robustness, particularly for older adult respondents.

Finally, some inconsistencies in user perceptions, such as differing views on navigation systems, highlight the diverse needs of assistive walker users. Our study aimed to capture a broad spectrum of experiences; however, these discrepancies suggest that additional segmentation or subgroup analysis may be beneficial in future research. Additionally, since our findings rely primarily on self-reported data, they may be influenced by subjective biases. Incorporating observational or experimental methods in future studies could provide a more comprehensive understanding of the real-world effectiveness and limitations of assistive walker designs.

\section{Conclusions}
This study establishes a foundation for designing improved assistive walkers tailored to the specific needs of older adults. The positive attitudes and knowledge among healthcare professionals and caregivers highlight the potential of these devices to enhance the quality of life for the elderly. By integrating advanced features such as GPS-enabled alarms, internal navigation systems, and grip force sensors, walkers can become safer, more user-friendly, and better address existing limitations.

This research emphasizes the importance of adaptability in walker design to meet the diverse needs of older adults. A user-centred approach, including real-world testing, is critical to creating devices that genuinely improve mobility and independence. Future research should explore integrating advanced technologies, ergonomic designs, and noise reduction features while increasing sample sizes by including independently living older adults. 

\begin{footnotesize}
    
\end{footnotesize}

\end{document}